\documentclass[galaxies,article,accept,moreauthors,pdftex,10pt,a4paper]{mdpi}
\usepackage[labelformat=simple]{subcaption}

\DeclareCaptionLabelFormat{subcaptionlabel}{\normalfont(\textbf{#2}\normalfont)}
\usepackage{latexsym}

\firstpage{1}
\articlenumber{x}
\doinum{10.3390/------}
\pubvolume{xx}
\pubyear{2018}
\copyrightyear{2018}
\history{Received: 19 January 2018; Accepted: 20 March 2018; Published: date}

\Title{The Train Wreck Cluster Abell 520 and the Bullet Cluster 1E0657-558 in a Generalized Theory of~Gravitation}

\Author{Norman Israel $^{1,}$* and John Moffat $^{2,3}$}

\AuthorNames{Norman Israel, John Moffat}

\address{%
$^{1}$ \quad Department of Physics, School of Natural and Applied Sciences, University of Technology, Jamaica, \mbox{237 Old Hope Road}, Kingston 6, Jamaica \\
$^{2}$ \quad Perimeter Institute for Theoretical Physics, Waterloo, ON N2L 2Y5, Canada; jmoffat@perimeterinstitute.ca\\
$^{3}$ \quad Department of Physics and Astronomy, University of Waterloo, Waterloo, ON N2L 3G1, Canada}

\corres{Correspondence: niquark@gmail.com}

\abstract{A major hurdle for modified gravity theories is to explain the dynamics of galaxy clusters. A case is made for a generalized gravitational theory called Scalar-Tensor-Vector-Gravity (STVG) or MOG (Modified Gravity) to explain merging cluster dynamics. The paper presents the results of a re-analysis of the Bullet Cluster, as well as an analysis of the Train Wreck Cluster in the weak gravitational field limit without dark matter. The King-$\beta$ model is used to fit the X-ray data of both clusters, and the $\kappa$-maps are computed using the parameters of this fit. The amount of galaxies in the clusters is estimated by subtracting the predicted $\kappa$-map from the $\kappa$-map data. The estimate for the Bullet Cluster is that $14.1\%$ of the cluster is composed of galaxies. For the Train Wreck Cluster, if the Jee et al. data are used, $25.7\%$ of the cluster is composed of galaxies. The baryon matter in the galaxies and the enhanced strength of gravitation in MOG shift the lensing peaks, making them offset from the gas. The work demonstrates that this generalized gravitational theory can explain merging cluster dynamics without dark matter.}


\begin{document}

\section{Introduction}
\label{intro}
In the past few decades, observations of galactic rotation speeds~\cite{VRubin,mpersic}, cluster mergers via gravitational lensing~\cite{cetal} and the cosmic microwave background~\cite{planck} have revealed gravitational anomalies that are interpreted by many astronomers as evidence that $26\%$ of the mass in the universe is unseen. Many expect that the existence of this missing mass (dark matter) will be confirmed by the discovery of new particles, which may be interacting or non-interacting~\cite{gbertone}. There are various candidates for this. However, at this time, experiments have turned up empty~\cite{LUX,PANDAXII,LC}. The interpretation of the anomalies as a problem of missing mass is based on the assumption that general relativity is applicable to large-scale structures such as galaxies and clusters. While general relativity is very well checked at the solar system scale, there is no reason to believe that it should survive large-scale structure tests unscathed. With~this in mind, with no dark matter particle candidate yet detected and the effects only observed to be gravitational, the authors see it justified to investigate a possible modification of general relativity that does not involve the addition of any extra mass. In either approach, at least one extra degree of freedom is added. The various approaches can be broken down to adopt the following strategies:
\begin{enumerate}[leftmargin=*,labelsep=5mm]
 \item Add extra mass in the regions where gravity is strongest. The mass could be:
 	\begin{enumerate}[leftmargin=2.3em,labelsep=4mm]
 	\item Non-interacting (this is the case for cold dark matter), \vspace{2pt}

 	\item Interacting (this creates a much broader class called the dark sector),
 	\end{enumerate}
 \item Add extra fields that implement various mechanisms.
\end{enumerate}

Some have chosen to call the extra degrees of freedom dark matter, regardless of whether they take the form of (1) or (2). However, relativistic modifications of gravity fall into the second category and implement very different strategies from those that fall into the first category. As a result, in this paper, the authors have only decided to call Category (1), dark matter (i.e., any theory that implements the strategy of (1) is, to the authors, a theory that contains dark matter, whereas those that do not fall into Category (2) and are said to possess no dark matter). A few ideas have been developed to take the second approach~\cite{mmilgrom,jbekenstein}. Some of these ideas fall under what are called mimetic gravity theories \cite{LG, AC, AG, AB, ACM, CM, OM, ND, DM}.

A major obstacle for many modified gravity theories is explaining cluster dynamics. This begs the question, does there exist a modification of general relativity that can address cluster dynamics without the addition of extra mass (dark matter)? In this paper, we will demonstrate that the Modified Gravity (MOG) theory called Scalar-Tensor-Vector-Gravity (STVG)~\cite{Moffat1} has the potential to explain merging galaxy cluster dynamics, thus answering the question in the affirmative. The theory is consistent with the LIGO GW170817 results \cite{GW170817} that gravitational waves and electromagnetic waves travel at the same speed \cite{mgreen}. Recall that there are three key areas where the anomalies show up:
\begin{itemize}[leftmargin=*,labelsep=5.8mm]
\item Galactic rotation curves
\item The dynamics of galaxy clusters
\item The cosmic microwave background
\end{itemize}

Any modification of general relativity that is intended to solve the `missing mass' anomaly without dark matter must address these three key areas where the anomalies show up. An earlier modified gravitational theory has already been shown to resolve anomalies of galaxy rotation speed~\cite{Moffat2,BrownsteinPhD,BrownsteinMoffat3}, and STVG has succeeded in this realm, as well.

Two key aspects of STVG are the following:
\begin{enumerate}[leftmargin=*,labelsep=5mm]
\item Gravity is stronger than is depicted in GR.
\item A repulsive gravitational force mediated by a spin-one vector field ($\phi$) screens gravity, making it appear weak at small distance scales such as the solar system and binary pulsars.
\end{enumerate}

In addition to its success in explaining galaxy rotation curves, STVG has shown success in cosmology. In STVG cosmology, there is an era where $\phi$ becomes dominant. This era is then followed by the later dominance of matter and the later enhancement of $G$, which in turn is followed by an accelerated era driven by the cosmological constant $\Lambda$. This reproduces the acoustic power spectrum, as well as the matter power spectrum. STVG cosmology is explained further in \cite{jmf, jamali, roshjam, jaif}.

This paper is focused on cluster dynamics. The study uses the Bullet Cluster and Train Wreck Cluster as test cases for this. In STVG, (1) is noticeable at galactic and galaxy cluster distance scales, while (2) guarantees agreement with the solar system and binary pulsar systems and is built into the Newtonian acceleration law. Recall that dark matter models assume general relativity is fine as it is and adds extra mass to produce the extra gravity needed to explain the galaxy rotation curves and galaxy cluster dynamics. As the gravity in STVG is sourced only by ordinary matter, it is considered an alternative to dark matter models. The key modification to the general relativity theory that leads to STVG is a spin-one massive particle called a ``phion''. The phion field interacts with ordinary matter with gravitational strength, and the fits of the theory to rotation curves and cluster dynamics yield a mass for the phion of $m_\phi=2.6\times 10^{-28}$ eV.

\begin{figure}[H]
	\centering
	\includegraphics[width=0.7\columnwidth]{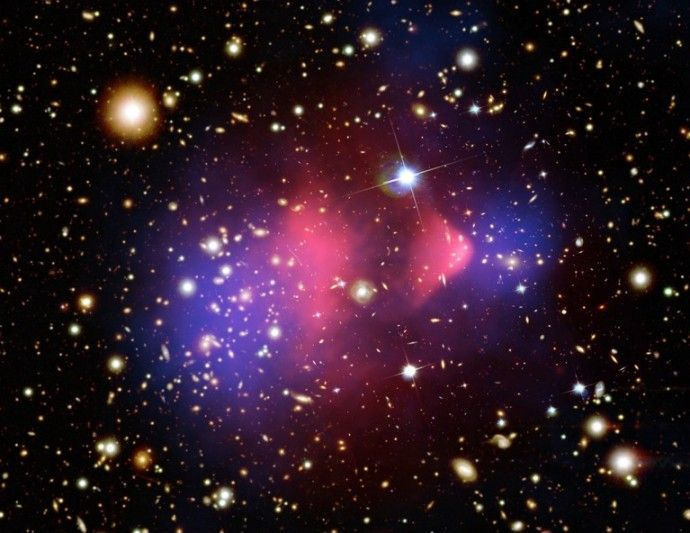}
	 \caption{$\kappa$-map (blue) superimposed on the gas (pink) for the Bullet Cluster. The blue regions are the regions with the largest amount of gravity and are where dark matter is said to be concentrated when viewed in the Einsteinian paradigm; the red region is the visible mass obtained from X-ray observations. This image is a Chandra X-ray Observatory image.} 
	\label{BCc}
\end{figure}

In 2006, a paper by Clowe et al.~\cite{Clowe} claimed that they had discovered direct evidence for dark matter through a cluster merger known as the Bullet Cluster (BC). The Bullet Cluster is a merger of two clusters where the gases in both clusters interact as the merger occurs. The gas slows down and heats up while emitting X-rays. The galaxies do not interact much, moving to the sides of the gas. As~a result, one ends up with a system that has the gas in the middle and the galaxies off to the sides. Figure~\ref{BCc}~\cite{Clowe} shows the lensing map (blue) superimposed on the gas (red). Notice the offset of the lensing map from the X-ray gas. It is this offset that resulted in the authors claiming a direct detection of dark matter, as one expects the lensing map to be on the gas and not offset from it. The common explanation for this in the context of dark matter is that there is non-interacting (cold) dark matter present in the sub-cluster (pink shock wave front on the right of Figure~\ref{BCc}) and the main cluster (pink region on the left of Figure~\ref{BCc}). When the merger occurred, the dark matter from the main and sub-clusters passed through each other moving off to the sides (blue regions of Figure~\ref{BCc}) and centred on the galaxies. While the results are not being questioned by the authors, the dark matter explanation is. The authors will show in this paper that there is an explanation in the context of STVG without postulating that extra exotic mass is sourcing the gravity. Rather, the apparent extra gravity is due to the significant decrease in the shielding effect of the repulsive vector field relative to the increase of the strength of gravity for the sizes of clusters. In addition, the offsets of the lensing peaks are due to the baryon matter in the galaxies shifting them away from the gas.

\begin{figure}[H]
	\centering
	\begin{subfigure}[b]{0.45\textwidth}
	\includegraphics[width=\textwidth]{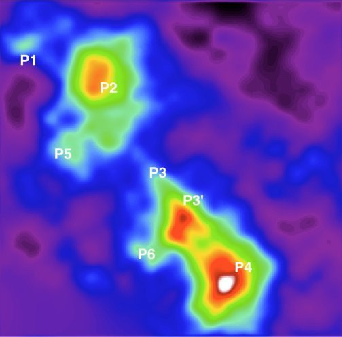}
	\caption{$\kappa$ map data}
	\label{md}
	\end{subfigure}
	\begin{subfigure}[b]{0.45\textwidth}
	\includegraphics[width=\columnwidth]{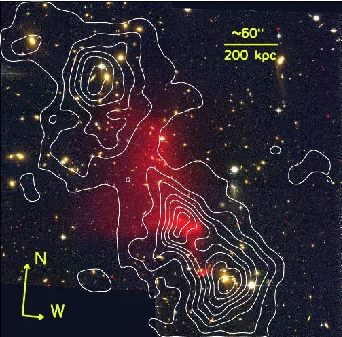}
	\caption{$\kappa$ map contours on gas}
	\label{sup}
	\end{subfigure}\vspace{-12pt}

	 \caption{Abell 520 (Train Wreck Cluster) $\kappa$-map. One peak (P3$'$) coincides with the gas (red)} 
	\label{A520_kpcont}
\end{figure}

In a later development, Mahdavi et al.~\cite{mahdetal} analysed the cluster merger known as Abell 520 (Train Wreck Cluster), claiming a lensing distribution, which challenges the notion that cold dark matter is all that is needed to explain clusters. We give in Figure~\ref{A520_kpcont} some plots from a 2014 study done by \mbox{Jee et al. \cite{JMetal}}. Figure~\ref{A520_kpcont}b shows the contours for the $\kappa$-map superimposed over the photon map (red). Figure~\ref{A520_kpcont}a shows the lensing map from Jee et al. There is a total of six lensing peaks, one of which (P3$'$) coincides with the gas. This is not expected in a merging system if one thinks of the system as dominated by cold dark matter, as cold dark matter (being non-interacting) is expected to produce a lensing peak offset from the gas in all cases. In the present study, we will explain all six peaks. The~presence of a peak coinciding with the gas was reaffirmed by a study performed by Jee et al.~\cite{jeetal}. Clowe et al.~\cite{cetal} disputed the existence of this peak, but a more recent detailed study by Jee et al.~\cite{JMetal} comparing the observational data used by Clowe et al. with theirs reaffirmed the existence of the significant lensing peak at the gas location. Such a central peak is naturally incorporated in STVG. The~peaks that are offset from the gas are again due to the galaxies containing baryons shifting them away from the gas.

Finally, a paper by Harvey et al.~\cite{Hetal}, which placed further constraints on a dark matter interacting cross-section, included in its study the Abell 520 cluster. This paper did not contain a central dark peak. This seems to leave the issue regarding the existence of this peak unresolved. The task of this paper is not to resolve the controversy of whether or not the central peak exists, but to rather make a preliminary attempt at demonstrating that it is possible for Scalar Tensor Vector Gravity (STVG) to explain this system if the peak is present.

We start with the weak field approximation of the STVG field equations and use the King-$\beta$ model to fit the X-ray data. We then use the parameters of the fit to compute the $\kappa$-map for both systems. Brownstein and Moffat have already done an analysis~\cite{BrownsteinMoffat2} of the bullet cluster using a previous version of a modified gravitational theory~\cite{Moffat2}, showing that the theory can explain the Bullet Cluster data without exotic dark matter. However, this study appears to have lacked a clear explanation of the offset of the lensing peaks from the gas. We performed a reanalysis of the Clowe et al. data using the more recent STVG theory~\cite{Moffat1,MoffatRahvar1} and attempted to suggest a clear explanation of the offset of the lensing peaks. The results are consistent with the results of the Brownstein--Moffat study, and the suggested offset of the peaks is that the baryons in the galaxies shifted the peaks from the gas as they moved to the sides. We believe that this study therefore shows that STVG provides a viable alternative to the cold dark matter explanation.

In addition to reaffirming the Bullet Cluster results, the authors investigated the cluster merger Abell 520 in the context of STVG. The Abell 520 study uses lensing data from the Jee et al. studies~\cite{JMetal}. We show that the lensing peak centred on the gas that has defied a cold dark matter explanation is~easily incorporated in STVG. As with the Bullet Cluster, the peaks that are offset from the gas are due to the galaxies that moved to the sides.

\section{Scalar-Tensor-Vector-Gravity}\label{STVG}

STVG (usually termed Modified Gravity (MOG)) is a modification of general relativity with a~massive vector field $\phi_{\mu}$, corresponding to a repulsive gravitational force with charge $Q_g$. The addition of this vector field leads to a modification of the law of gravitation beyond solar system scales. A~central feature of the theory is the dynamical gravitational coupling $G$. This coupling varies both spatially and temporally. In addition, the theory introduces the scalar field $\mu$, which is the mass of the vector field. The theory has two (2) degrees of freedom: $\alpha$ and $\mu$.

\subsection{STVG Field Equations}

The MOG action is given by:
\begin{align}
S=S_G+S_\phi+S_S+S_M,
\end{align}
where $S_M$ is the matter action and:
\begin{align}
S_G&=\frac{1}{16\pi}\int d^4x\sqrt{-g}\left[\frac{1}{G}(R+2\Lambda)\right],\\
S_\phi&=\int d^4x\sqrt{-g}\left[-\frac{1}{4}B^{\mu\nu}B_{\mu\nu}+\frac{1}{2}\mu^2\phi^\mu\phi_\mu\right],\\
S_S&=\int d^4x\sqrt{-g}\left[\frac{1}{G^3}\left(\frac{1}{2}g^{\mu\nu}\partial_\mu G\partial_\nu G-V_G\right)+\frac{1}{\mu^2G}\left(\frac{1}{2}g^{\mu\nu}\partial_\mu\mu\partial_\nu\mu-V_\mu\right)\right],
\end{align}
where
$B_{\mu\nu}=\partial_\mu\phi_\nu-\partial_\nu\phi_\mu$ and $V_G$ and $V_\mu$ are potentials. Note that we choose units such that
$c=1$, using the metric signature $[+,-,-,-]$. The Ricci tensor is:
\begin{equation}
R_{\mu\nu}=\partial_\lambda\Gamma^\lambda_{\mu\nu}-\partial_\nu\Gamma^\lambda_{\mu\lambda}
+\Gamma^\lambda_{\mu\nu}\Gamma^\sigma_{\lambda\sigma}-\Gamma^\sigma_{\mu\lambda}\Gamma^\lambda_{\nu\sigma}.
\end{equation}

The matter stress-energy tensor is obtained by varying the
matter action $S_M$ with respect to the~metric:
\begin{align}
 T^{\mu\nu}_M=-2(-g)^{-1/2}\delta S_M/\delta g_{\mu\nu}\,.
\end{align}

Varying $S_\phi+S_S$ with respect to the metric yields:
\begin{align}
 T^{\mu\nu}_{\rm MOG}=-2(-g)^{-1/2}[\delta S_\phi/\delta
 g_{\mu\nu}+\delta S_S/\delta g_{\mu\nu}]\,.
\end{align}

Combining these gives the total stress-energy tensor:
\begin{align}
 T^{\mu\nu}=T^{\mu\nu}_{\rm M}+T^{\mu\nu}_{\rm MOG}\,.
\end{align}

The MOG field equations are given by:
\begin{align}
 G_{\mu\nu}-\Lambda g_{\mu\nu}+Q_{\mu\nu}=8\pi GT_{\mu\nu}\,,
 \label{eq:MOGE}
\end{align}
\begin{equation}
\frac{1}{\sqrt{-g}}\partial_\mu\biggl(\sqrt{-g}B^{\mu\nu}\biggr)+\mu^2\phi^\nu=-J^\nu,
\end{equation}
\begin{equation}
\partial_\sigma B_{\mu\nu}+\partial_\mu B_{\nu\sigma}+\partial_\nu B_{\sigma\mu}=0.
\end{equation}

Here, $G_{\mu\nu}=R_{\mu\nu}-\frac{1}{2} g_{\mu\nu} R$
is the Einstein tensor and:
\begin{align}
 Q_{\mu\nu}=\frac{2}{G^2}(\partial^\alpha G \partial_\alpha G\,g_{\mu\nu}
 - \partial_\mu G\partial_\nu G) - \frac{1}{G}(\Box G\,g_{\mu\nu}
 - \nabla_\mu\partial_\nu G)
 \label{eq:Q}
\end{align}
is a term resulting from the the presence of second derivatives
of $g_{\mu\nu}$ in $R$ in $S_G$. We also have the field~equations:
\begin{equation}
\Box G=K,\quad \Box\mu=L,
\end{equation}
where $\Box=\nabla^\mu\nabla_\mu$, $K=K(G,\mu,\phi_\mu)$ and $L=L(G,\mu,\phi_\mu)$.

Combining the Bianchi identities, $\nabla_\nu G^{\mu\nu}=0$, with the
field Equation (\ref{eq:MOGE}) yields the conservation~law:
\begin{align}
 \nabla_\nu T^{\mu\nu}+\frac{1}{G}\partial_\nu G\,T^{\mu\nu} -
 \frac{1}{8\pi G}\nabla_\nu Q^{\mu\nu}=0 \,.
 \label{eq:Conservation}
\end{align}

It is a key premise of MOG that all baryonic matter possesses, in
proportion to its mass $M$, positive gravitational charge:
$Q_g=\kappa_g\,M$. This charge serves as the source of the vector field
$\phi^\mu$. Moreover, $\kappa_g=\sqrt{G-G_N}=\sqrt{\alpha\,G_N}$, where
$G_N$ is Newton's gravitational constant and
\mbox{$\alpha=(G-G_N)/G_N\ge 0$.} Variation of $S_M$ with respect to the
vector field $\phi^\mu$ then yields the MOG four-current
$J_\mu=-(-g)^{-1/2}\delta S_M/\delta \phi^\mu$.

For the case of a perfect fluid:
\begin{equation}
\label{energymom}
T^{\mu\nu}_M=(\rho+p)u^\mu u^\nu-pg^{\mu\nu},
\end{equation}
where $\rho$ and $p$ are the matter density and pressure, respectively, and $u^\mu$ is the four-velocity of an element of the fluid.
We obtain from (\ref{energymom}) and $u^\mu u_\mu=1$ the four-current:
\begin{equation}
J_\mu=\kappa_g T_{M\mu\nu}u^\nu=\kappa_g\rho u_\mu.
\end{equation}

It is shown in~\cite{mrosh} that, with the assumption $\nabla_\mu J^\mu=0$,
(\ref{eq:Conservation}) reduces to:
\begin{align}
\nabla_\nu T_M^{\mu\nu}=B_\nu^{~\mu} J^\nu\,.\label{eq:Mcons}
\end{align}
\subsection{STVG Weak Field Approximation}

In the weak field approximation, we perturb $g_{\mu\nu}$ about the Minkowski metric $\eta_{\mu\nu}$:
\begin{equation}
g_{\mu \nu} = \eta_{\mu \nu} + \epsilon h_{\mu \nu}.
\end{equation}

The condition $g_{\mu \nu}g^{\mu \rho} = \delta_{\nu}^{~\rho}$ demands that $g^{\mu \rho} = \eta^{\mu \rho} - \epsilon h^{\mu \rho}$.

The field equation for the vector field $\phi^{\mu}$ is given by:
\begin{equation}
\partial_\nu B^{\mu \nu} + \mu^2 \phi^{\mu} = - J^{\mu}.
\label{current}
\end{equation}

Assuming that the current $J^{\mu}$ is conserved, $\partial_\mu J^{\mu} = 0$, then in the weak field approximation, we can
apply the condition $\partial_{\mu}\phi^{\mu} = 0$. In the static limit, one obtains for the source-free spherically symmetric~equation:
\begin{equation}
\nabla^2\phi^{0} - \mu^2\phi^0 = 0,
\end{equation}
which has the point particle solution:
\begin{equation}
\label{phisolution}
\phi_0(r)=-Q_g\frac{\exp(-\mu r)}{r},
\end{equation}
where the charge $Q_g=\sqrt{\alpha G_N}M=\kappa_g M$. This solution can be written for a distribution of matter as:
\begin{equation}
\phi_0 = -\int \frac{e^{-\mu |\vec{x} - \vec{x}'|}}{|\vec{x} - \vec{x}'|}J_0(\vec{x}')d^3x',
\end{equation}
where:
\begin{equation}
Q_g=\int J_0d^3x.
\end{equation}

We assume that in the slow motion and weak field approximation $dr/ds\sim dr/dt$ and $2GM/r\ll 1$, then for the radial
acceleration of a test particle, we get:
\begin{equation}
\frac{d^2r}{dt^2}+\frac{GM}{r^2}=\frac{q_gQ_g}{m}\frac{\exp(-\mu r)}{r^2}(1+\mu r),
\end{equation}
where $q_g=\kappa_g m$. For the static solution (\ref{phisolution}) $q_gQ_g/m=\alpha G_NM$, and $G=G_N(1+\alpha)$, which
yields the acceleration law:
\begin{equation}
\label{accelerationlaw}
a(r)=-\frac{G_NM}{r^2}\biggl\{1+\alpha\biggl[1-\exp(-r/r_0)\biggl(1+\frac{r}{r_0}\biggr)\biggr]\biggr\},
\end{equation}
where $r_0=1/\mu$. We observe that the acceleration of a particle is independent of its material content (weak equivalence principle). The acceleration law can be extended to a distribution of matter: 
\begin{multline}
\label{accelerationlaw2}
a({\vec x})=-G_N\int d^3{\vec x}'\frac{\rho({\vec x}')({\vec x}-{\vec x}')}{|{\vec x}-{\vec x}'|^3}
[1+\alpha-\alpha e^{-\mu |\vec{x}-\vec{x}'|}(1+\mu |\vec{x}-\vec{x}'|)].
\end{multline}

We can write (\ref{accelerationlaw2}) as:
\begin{equation}
a({\vec x})=- \int G({\vec x}-{\vec x}')\frac{\rho({\vec x}')({\vec x}-{\vec x}')}{|{\vec x}-{\vec x}'|^3}d^3x',
\end{equation}
where the effective gravitational coupling strength is given by:
\begin{equation}
G({\vec x}-{\vec x}') = G_N[1+\alpha-\alpha e^{-\mu |\vec{x}-\vec{x}'|}(1+\mu |\vec{x}-\vec{x}'|)].
\label{runG}
\end{equation}

From Equation~\eqref{runG}, we see that $G$ is not, in general, constant. However, for sufficiently large systems, such as galaxy clusters, it turns out that $G$ behaves as if it is constant. The~STVG fits to galaxy rotation curves, and galaxy cluster dynamics have yielded the best fit values\mbox{ $\alpha=8.89\pm 0.34$} and \mbox{$\mu=0.042\pm 0.004~{\rm kpc}^{-1}$~\cite{MoffatRahvar1,MoffatRahvar2}}. The value for $\mu$ corresponds to the vector field mass \mbox{$m_\phi=2.6\times 10^{-28}$ eV}. The Tully--Fisher law relating the galaxy luminosities to the flat rotation curves is deduced from the STVG dynamics in excellent agreement with the data~\cite{MoffatRahvar1}. In the fits to the dynamics of galaxy clusters~\cite{MoffatRahvar1}, the Yukawa repulsive exponential term produced by the vector field $\phi_\mu$ does not have a significant effect when one uses $\mu^{-1}=24~\textrm{kpc}$. As a result, one is only left with $G = G_N(1+\alpha)$, which is used to compute the $\kappa$-convergence lensing map for the Bullet Cluster and Train Wreck Cluster.

\subsection{Gravitational Lensing in STVG}

From \cite{Peacock}, the Poisson equation with the lensing potential gives:
\begin{equation}
\nabla^2_{\theta}\psi = \frac{D_L D_{LS}}{D_S}\frac{8\pi G_N}{c^2}\Sigma = 2\frac{\Sigma}{\Sigma_c},
\end{equation}
where:
\begin{equation}
\frac{\Sigma(x,y)}{\Sigma_c} = \kappa(x,y).
\end{equation}
and:
\begin{equation}
\Sigma(x,y) = \int^{z_{\textrm{out}}}_{-z_{\textrm{out}}}\rho(x,y,z)dz.
\label{2d_density}
\end{equation}
with $z_{\textrm{out}} = \sqrt{r^2_{\textrm{out}} - x^2 - y^2}$, with:
\begin{equation}
r_{\textrm{out}} = r_c\left[\left(\frac{\rho_{0}}{10^{-28}\textrm{g}/\textrm{cm}^3}\right)^{2/(3\beta)} - 1\right]^{1/2},
\end{equation}
where $10^{-28}~\textrm{g}/\textrm{cm}^3$ is the total density of the cluster at $r_{\textrm{out}}$.
One has:
\begin{equation}
\Sigma(x,y) = \sum_{i=1}^{n}\int_{-z_{\textrm{out},i}}^{z_{\textrm{out},i}}\rho_i(0)\left[1+\frac{x_i^2+y_i^2+z_i^2}{r^2_{ci}}\right]^{-3\beta_i/2}dz
\label{dens0}
\end{equation}
where $i$ is the index representing the i
-th cluster, $\Sigma_c \approx 3.1\times 10^9$ {$\textrm{M}_{\odot}\textrm{kpc}^{-2}$} for the Bullet Cluster and $\Sigma_c \approx 3.8\times 10^9~\textrm{M}_{\odot}\textrm{kpc}^{-2}$ for the Train Wreck Cluster. $\kappa$ is a measure of the curvature of space-time. In~STVG, we have the surface density defined as \citep{BrownsteinMoffat2}: 
\begin{equation}
\bar{\Sigma}(x,y) = (1+\alpha)\int\rho (x,y,z)dz,
\label{sbar}
\end{equation}
so that for STVG:
\begin{equation}
\phi (x,y) = \frac{\bar{\Sigma}(x,y)}{\Sigma_c}.
\label{k_map_stvg}
\end{equation}

In this paper, Equation~\eqref{k_map_stvg} is first computed. This is not the entire $\kappa$ map, as this only considers the contribution of the gas. In order to get the entire $\kappa$ map, one must also consider the contribution of the galaxies:
\begin{equation}
\kappa(x,y) = \frac{\bar{\Sigma}(x,y)+\bar{\Sigma}_{\textrm{gal}}(x,y)}{\Sigma_c}
\label{subtgal}
\end{equation}

When we solve (\ref{subtgal}) for the galaxies, we get:
\begin{equation}
\Sigma_{\textrm{gal}}(x,y)\approx \frac{\kappa(x,y)\Sigma_c - \bar{\Sigma}(x,y)}{(1+\alpha)}.
\label{galsub}
\end{equation}

We then compute the mass of the galaxies by integrating over $\Sigma_{\textrm{gal}}(x,y)$:
\begin{equation}
M_{\textrm{gal}} = \int \int \Sigma_{\textrm{gal}}(x,y)dxdy
\label{mgal}
\end{equation}

In the next few sections, the results of the Bullet Cluster and Train Wreck Cluster studies will be discussed. These studies were done by starting with the King-$\beta$ model. We will discuss this first. The~$\kappa$-map computed in both cases using STVG is shown to be consistent with the Bullet Cluster and Train Wreck Cluster data.

\section{The King-\boldmath{$\beta$} Model}
\label{king}

In the following sections, the recent work done to check the results of~\cite{BrownsteinMoffat2} will be discussed, as well as work done by~\cite{mahdetal}. This study was done first by assuming an almost isothermal gas sphere for the Bullet Cluster and Train Wreck Cluster. With this assumption, we resort to the King-$\beta$ model~\cite{king}. We thus have:
\begin{equation}
\rho(r) = \sum_{i=1}^{n}\rho_i(0)\left[1+\left(\frac{r_i}{r_{ci}}\right)^2\right]^{-3\beta_i/2},
\label{dens}
\end{equation}
where $\rho(0)$ is the central density for a given X-ray peak and $r_i$ is the radial distance for that peak. Using~the fact that galaxy clusters have a finite spatial extent, one can connect the King-$\beta$ model with data by integrating Equation~(\ref{dens}) along the line-of-sight, giving the total surface density (\ref{2d_density}):
\begin{equation}
\Sigma(x,y) = \sum_{i=1}^{n}\Sigma_i(0)\left[1+\frac{x_i^2+y_i^2}{r^2_{ci}}\right]^{-(3\beta_i-1)/2},
\label{siggp}
\end{equation}
where $\Sigma_i(0)$ is the central surface mass density for a particular peak. We assume:
\begin{equation}
	X = \begin{bmatrix}
		x \\
		y \\
		z
		\end{bmatrix}
\end{equation}

The centre of the $i$-th cluster is $X_{0,i}$ and $X_i = X - X_{0,i}$, and $r_i = \sqrt{X^2_{i}}$. A derivation of Equation~\eqref{siggp} is given in Appendix \ref{sigmap}.

We fit Equation~\eqref{siggp} to the $\Sigma$-map for the Bullet Cluster data. The parameters are then used to compute the $\kappa$-map. In cases where the geometry is complicated such as with the Train Wreck Cluster, we fit the X-ray peaks with surface brightness profiles given by:
\begin{equation}
I(r) = \sum_{i=1}^{n} I_i(0)\left[1+\left(\frac{r_i}{r_{ci}}\right)^2\right]^{-3\beta_i+1/2},
\label{sbright}
\end{equation}
where $I_i(0)$ is the central brightness for a given X-ray peak. The parameters of this fit are used to compute the $\kappa$-map of the system.

\section{Results and Analysis}\label{results}

In this section, we will discuss in detail the studies done on both the Train Wreck Cluster and Bullet Cluster in the context of STVG. We first start by discussing the results of the bullet cluster, fitting the lensing peaks only using the gas from the main cluster and sub-cluster, as well as the galaxies to the sides.

Finally, we discuss work done on the Train-Wreck Cluster. We focus on the central lensing peak (centred on the gas) given by~\cite{JMetal}. Recall that the results from~\cite{JMetal} show that there is a larger than expected lensing peak at the central gas, whereas other studies, such as~\cite{Hetal}, do not show this peak. Our intent is not to resolve this controversy, but rather to demonstrate that the central peak is naturally incorporated into STVG.

Recall that the Yukawa repulsive exponential term produced by the vector field $\phi_\mu$ does not have a significant effect when one uses $\mu^{-1}=24~\textrm{kpc}$. As a result, one is only left with $G = G_N(1+\alpha)$, which is used to compute the $\kappa$-convergence lensing map for the Bullet Cluster and the Train Wreck~Cluster.

\subsection{Analysis of Bullet Cluster Using STVG}\label{analysis}

The Bullet Cluster is a merging system located at $z=0.296$ in the constellation Carina. It was first studied in 2006 by Clowe et al.~\cite{Clowe}. The system is a dramatic demonstration of a gravitational anomaly where a lensing peak is clearly separate from the visible gas. This distinct separation of the visible matter from the lensing peaks is considered by many as evidence that the lensing peaks need to be sourced with extra mass even in the case of a modified gravity theory. In this section, we will show that this system is naturally explained by a modified gravity theory without the need for the addition of extra mass.

In the following:
\begin{enumerate}[leftmargin=*,labelsep=5mm]
\item We fit the main and subcluster (simultaneously) of the $\Sigma$-map with (\ref{siggp}), using a least squares fitting routine. This describes a double-$\beta$ model. From the fitting, we obtain the parameters $\beta$, \mbox{$\Sigma_0$ and $r_c$}.
\item We then use the parameters from the double-$\beta$ model to compute the galaxy amount using (\ref{sbar}) and (\ref{k_map_stvg}) to obtain a reasonable $\alpha$.
\item Using this $\alpha$, we compute the $\kappa$-map.
\end{enumerate}

We compare the result with that of Brownstein--Moffat~\cite{BrownsteinMoffat2}. We find that the amount of galaxies account for a mass of about $14.1\%$ using the double-$\beta$ model. This was obtained using an $\alpha = 2.32$. The result is only $3\%$ different from the estimate obtained in~\cite{BrownsteinMoffat2}, which gave a $17\%$ estimate (using an $\alpha = 2.82$). Relying on the $\beta$-model assumes spherical symmetry. We believe there is no issue with this for this study because in \cite{BrownsteinMoffat2}, it was demonstrated that this assumption leads to a very good prediction of the temperature of the main cluster of the BC. They obtained a result of $15.5\pm 3.9$ keV, which is very good when compared with the value of $14.8^{+2.0}_{-1.7}$ keV obtained empirically. Their calculated value was then successfully used to reproduce the mass profile of the main cluster ICM.

\subsubsection{Assumptions}

In this study, we made the following assumptions:

\begin{enumerate}[leftmargin=*,labelsep=5mm]
\item The gas has an isothermal distribution and can be modelled via the King-$\beta$ model (from Section~\ref{king}).
\item All the visible matter in the system consists only of the X-ray-emitting gas and the galaxies.
\end{enumerate}

The first assumption allows a means of fitting the X-ray data ($\Sigma$-map) with the King-$\beta$ model, while the second assumption allows us to compute the $\kappa$-map from the X-ray data only. The galaxies are then taken to be sufficient to explain the remaining gravity, as well as the offset of the lensing peaks from the gas. We emphasise that the first assumption should be valid for the main gas plasma, for the assumption of gas equilibrium (with the exception of the shock wave part) allows for the above stated very good prediction of the temperature of the main cluster in agreement with the observational data.

\subsubsection{The $\Sigma$-Map for the Bullet Cluster}

\begin{figure}[H]
	\centering
	\begin{subfigure}[b]{0.48\textwidth}
	\centering 
	\includegraphics[width=\textwidth]{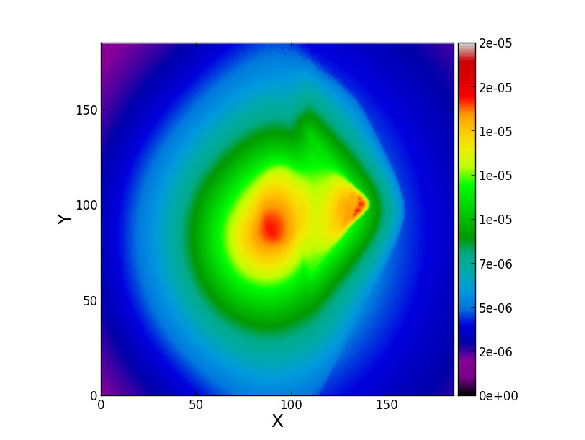}
	\caption{$\Sigma$ map data}
	\label{bcsmp}
	\end{subfigure}
	\begin{subfigure}[b]{0.48\textwidth}
	\centering 
	\includegraphics[width=\textwidth]{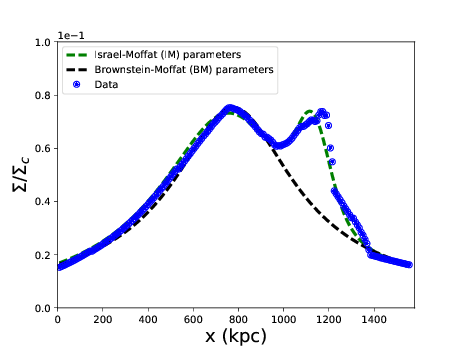}

	\caption{STVG prediction}
	\label{bcsmp2d}
	\end{subfigure}	\vspace{-12pt}
	\caption{Surface density map of the Bullet Cluster (BC) ({left}) and 2D projection of the surface density plot ({right}). The~subcluster is the shock wave front on the right (in top plot), while the main cluster is on the left (top plot). The colour scale corresponds to the mass in units of $10^{15}~\textrm{M}_{\odot}$ and has a resolution of $8.5~\textrm{kpc}~\textrm{pixel}^{-1}$. This is based on a redshift measurement of the BC of $260~\textrm{kpc}$~{$\textrm{arcmin}$}$^{-1}$. In the 2D projection (bottom), the green dotted curve is the fit to the main and sub-cluster data (blue points) as a double-$\beta$ model. The black dotted curve is the single-$\beta$ model fit using the Brownstein--Moffat parameters. STVG, Scalar-Tensor-Vector-Gravity.} 
	\label{surf_den}
\end{figure}

The surface density map data for the Bullet Cluster is shown in Figure~\ref{surf_den}a. It is a 185 $\times$ 185 $\textrm{px}^2$ image with a mass resolution of $10^{15}~\textrm{M}_{\odot}/\textrm{px}^2$. The farthest region on the right is the sub-cluster, while the region on the left is the main-cluster. A 2D projection of Figure~\ref{surf_den}a with the fitting according to the King-$\beta$ model is shown in Figure~\ref{surf_den}b. It spans 185 pixels or 1572.5 kpc (using the fact that for the $\Sigma$-map, the distance resolution is about 8.5 kpc/px). The data are represented by the blue points. There~are two~peaks, the widest being the main-cluster, and the smallest is the sub-cluster. The black curve is the fit using the Brownstein--Moffat parameters, which are given for a single-$\beta$ model for the main cluster in Table~\ref{paraB}. The green curve is the fit using the double-$\beta$ model parameters obtained in this study. This~is shown in Table~\ref{paraB} for the double-$\beta$ model (IM
, main + sub-cluster).

\begin{table}[htb!]
\caption{Values of parameters obtained from the fitting in Figure~\ref{surf_den}b. The values of $\Sigma$ are multiples of $10^{15}~\textrm{M}_{\odot}/\textrm{px}^2$.}
\begin{center}
  \begin{tabular}{ccc }
  \toprule
  \textbf{Model Type} & \textbf{Property} & \textbf{Values} \\
  \midrule
  \multirow{3}{*}{Single $\beta$ (BM)}
  & $\beta_1$ & $0.803$ \\
  & $r_{c1}$ & $278.0\:\textrm{kpc}$ \\
  & $\Sigma_{01}$ & $1.6859\times 10^{-5}$ \\
 \midrule
  \multirow{6}{*}{Double $\beta$ (IM)}
  & $\beta_1$ & $0.91 \pm 0.03$ \\
  & $\beta_{2}$ & $3.54 \pm 1.45$ \\
  & $r_{c1}$ & $355.6 \pm 14.6\:\textrm{kpc}$ \\
  & $r_{c2}$ & $224.6 \pm 58.4\:\textrm{kpc}$ \\
  & $\Sigma_{01}$ & $(1.639 \pm 0.009)\times 10^{-5}$ \\
  & $\Sigma_{02}$ & $(0.781 \pm 0.162)\times 10^{-5}$ \\
  \bottomrule
  \end{tabular}
\end{center}
\label{paraB}
\end{table}

The parameters of Table~\ref{paraB} and Equation~\eqref{k_map_stvg} were used to compute the $\kappa$-map, which will be discussed~next.

\subsubsection{The Convergence $\kappa$-Map for the Bullet Cluster}

In order to compute the $\kappa$-map, recall that the following was done:
\begin{enumerate}[leftmargin=*,labelsep=5mm]
\item We do a galaxy subtraction to obtain a reasonable $\alpha$ using the parameters from the double-$\beta$ model.
\item We then use these parameters (including the obtained $\alpha$) and (\ref{sbar}) and (\ref{k_map_stvg}) to compute the $\kappa$-map.
\end{enumerate}

\begin{figure}[H]
\centering
	\hspace{2em}\begin{subfigure}[b]{0.4\textwidth}
	\centering
	\includegraphics[width=\textwidth]{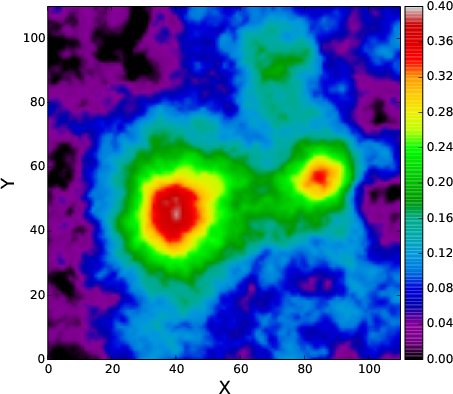}
	\caption{Bullet cluster $\kappa$-map data.}
	\label{kap_mp}
	\end{subfigure}
	\begin{subfigure}[b]{0.5\textwidth}
	\centering
	\includegraphics[width=\textwidth]{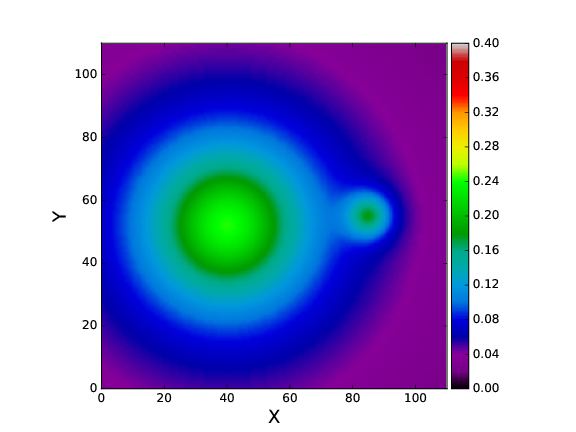}
	\caption{STVG prediction.}
	\label{mog_pd}
	\end{subfigure}
	\vspace{-12pt}

	 \caption{$\kappa$-map of the BC. The resolution is $15.4~\textrm{kpc}~\textrm{pixel}^{-1}$. This is based on a redshift measurement of the BC of $260~\textrm{kpc}~\textrm{arcmin}^{-1}$.} 
	\label{kap_kp}
\end{figure}

The data and STVG prediction are shown in Figure~\ref{kap_kp}a,b, respectively. The density maps are $110\times 110~\textrm{px}^2$.

Comparing the density plots of Figure~\ref{kap_kp}a,b shows that the gas alone cannot explain the lensing peaks fully. Following Assumption (2), the rest can only be explained by the galaxies.

\subsubsection{Including the Galaxies}

In cluster mergers, the gases interact while the galaxies move to the sides as they do not interact significantly. As a result, any system modelling cluster mergers must consider the effect of these galaxies and the baryonic matter they contain. In the modified gravity theory explored in this paper, it~is proposed that these galaxies are responsible for producing the offset of the lensing peak from the central gas. The baryon matter in the galaxies shifts the lensing peak from the baryon matter in the gas, producing the offset. The enhancing of the lensing peaks is due to the enhanced gravitational coupling $G=G_N(1+\alpha)$.

\begin{figure}[H]
	\centering
	\includegraphics[width=0.8\textwidth]{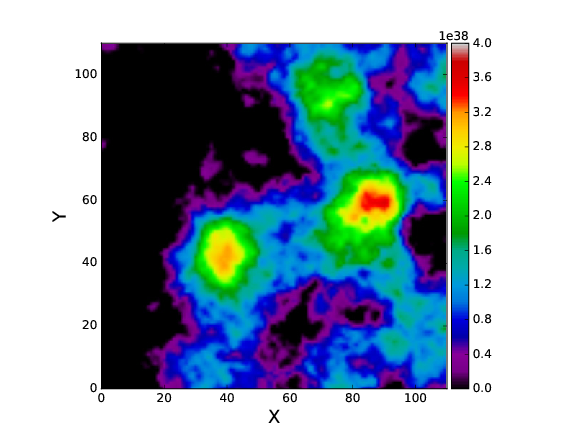}
	 \caption{{Galaxy subtraction} for the bullet cluster. There is galaxy contributions for both the main cluster and sub-cluster.}
	\label{BC_gal}
\end{figure}

In order to compute the galaxy mass, we use the parameters obtained from the $\Sigma$-map fitting. Figure~\ref{BC_gal} shows the result for the inclusion of the galaxies for the Bullet Cluster. The results of this computation yields a galaxy contribution of around $14.1\%$ using the double-$\beta$ model. This~result ignores all negative mass contributions as these are not physical. Including the negative mass contributions gives a galaxy contribution of about $10.2\%$. These negative mass contributions stem from the fact that the predictions come from a zeroth order approximation, which is much more uniform than the data. The result is that some regions will have a relatively higher $\kappa$ than the same region of the data. Recall that we are looking at the difference between the data and prediction; hence, 
the~resulting galaxy mass contribution for such regions is negative. The result when we ignore the negative mass contribution is close to the Brownstein--Moffat~\cite{BrownsteinMoffat2} estimate of a $17\%$ galaxy contribution. In the Brownstein--Moffat study, only the main cluster $\Sigma$-map was fitted. As a result, a major part of the $17\%$ of the galaxies would come from the sub-cluster. Brownstein and Moffat considered the model with $\alpha = \sqrt{\frac{M_0}{M(r)}}$, where $M_0$ is a MOG mass scale and $M(r)$ is the total baryonic mass enclosed in a spherical region of radius $r$~\cite{BrownsteinMoffat2}. The choice of this model can cause a difference in the final calculations of the galaxy mass and lensing.

Our results demonstrate that STVG can explain the Bullet Cluster without the need for dark matter. The mass of the visible matter (gas and galaxies) is sufficient to explain the size of the lensing peaks of the system and their offset from the gas.

\subsection{Analysis of the Train Wreck Cluster Using STVG}

Abell 520 is a cluster merger at $z=0.2$ in the constellation Orion. Although the Abell 520 merger also has lensing peaks that are clearly separated from the gas as with the Bullet Cluster, its situation is quite different from the Bullet Cluster, because of the complicated geometry and what appears to be a significant lensing peak centred on the gas (though the existence of this peak is being debated). This large lensing peak, if it exists, presents a problem for cold dark matter models. This is because the non-interacting nature of cold dark matter means it should move to the sides of the gas as the galaxies do. The presence of this peak has led to the proposal of a self-interacting dark matter explanation. This would imply that the system might have a mixture of interacting and non-interacting dark matter. In STVG, such a peak is not difficult to explain, for only the baryonic matter in the gas is needed to explain it.

To do the calculations for this cluster in the context of STVG, (\ref{sbright}) is fitted to the surface brightness data with $n=3$. The central brightness for each peak (which is a fitting parameter) is used to estimate a central surface density corresponding to that peak. This and the other fitting parameters are used to compute the galaxy contribution and estimate $\alpha$. The galaxy contribution computed is $25.7\%$ using an $\alpha = 34.20$. Using this $\alpha$ and the King-$\beta$ model parameters obtained, the $\kappa$-map was calculated.

\subsubsection{Assumptions}

In this study, we make the following assumptions:

\begin{enumerate}[leftmargin=*,labelsep=5mm]
\item The gas is in hydrostatic equilibrium and can be modelled by the King-$\beta$ model described in Section~\ref{king}.
\item All the visible matter in the system consists only of the X-ray-emitting gas and the galaxies.
\item The surface density of the gas within the region of size 150 kpc of P3$'$ is sufficient to estimate the central surface densities. The surface densities are related by:
\begin{equation}
\Sigma_{0i} = \frac{P_{0i}}{P'}\Sigma_{(r\leq 150 \textrm{kpc})}
\label{cent}
\end{equation}
\end{enumerate}
where $P_{0i}$ is the `photon count' for the i-th peak and $P'$ is obtained using the total upper limit estimate of the gas mass shown in Table 1 of \cite{JMetal}.

Using these assumptions, the X-ray data are fitted, and the parameters are then used to estimate $\alpha$ from a galaxy estimate, after which this $\alpha$ was used to compute the $\kappa$-map.

\subsubsection{The X-Ray Photon-Map for the Train Wreck Cluster}

\begin{figure}[H]
\centering
	\hspace{0.5em}\begin{subfigure}[b]{0.45\textwidth}
	\centering 
	\includegraphics[width=\textwidth]{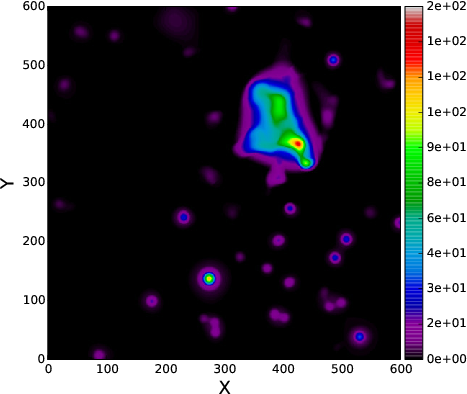}
	\caption{Smoothed X-ray data}
	\label{smpd}
	\end{subfigure}
	\begin{subfigure}[b]{0.53\textwidth}
	\centering 
	\vskip 1.75em
	\includegraphics[width=\textwidth]{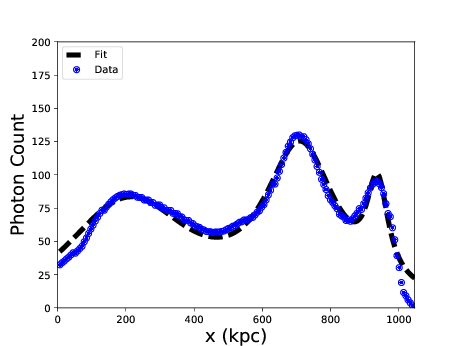}
	\caption{2D projection of X-ray data}
	\label{smp2d}
	\end{subfigure}
	\vspace{-12pt}

	 \caption{{A520 photon map}. This shows the smoothed photon map distribution (\textbf{a}) and the fitting of the 2D projection of the photon map distribution along the \emph{x}-axis (\textbf{b}) with the fitting curve (black dotted curve).}
	\label{A520_sfit}
\end{figure}

There is no gas mass data for the Train Wreck Cluster
, so, instead, photon spectra are used, which correspond to upper mass limits. The smooth X-ray photon spectra distribution data are shown in Figure~\ref{A520_sfit}a. It is a $600\times 600~\textrm{px}^2$ image with a distance resolution of $6.5~\textrm{kpc}/\textrm{px}$, based on a $200~\textrm{kpc}/\textrm{arcmin}$ measured redshift distance from \cite{jeetal}. In order to calculate the $\kappa$-map, the first step is to fit the photon spectra and use the parameters of the fit to compute the $\kappa$-map. Since the photon spectra are not a $\Sigma$-map, we use the result of~\cite{JMetal} and the total pixel area covered by the gas within 150 kpc. In addition, an upper limit estimate of the total mass of the gas obtained is used to compute the surface mass density of the gas within the 150-kpc region. We compute an upper limit to the surface density for that region ($\Sigma_{(r\leq 150 \textrm{kpc})}$). Moreover, we computed $\Sigma_{(r\leq 150 \textrm{kpc})} = 2.2\times 10^9~\textrm{M}_{\odot}/\textrm{px}^2$. Finally, we found $P'$ to be 2925.0 by seeing what value of $P'$ it takes to get the total upper limit gas mass estimate using Equation~\eqref{cent} when we sum over the entire $\Sigma$-map. We then computed the central surface density for each peak via \eqref{cent}.

The 2D projection of the photon spectra is shown in Figure~\ref{A520_sfit}b along with the fitting curve (black dotted curve). The parameters obtained from the fitting of the photon map in Figure~\ref{A520_sfit} are shown in Table~\ref{paran}. These are the $\beta_{0i}$'s, $r_{ci}$'s and $I_{0i}$'s. These parameters are used in the computation of the $\kappa$-map. In the next few sections, we will discuss the computation of the $\kappa$-map.

\vspace{-12pt}

\begin{table}[H]
\centering
\caption{Values of parameters obtained from the fitting in Figure~\ref{surf_den}b.}

  \begin{tabular}{ ccc}
  \toprule
  \textbf{Model Type }& \textbf{Property }& \textbf{Values }\\
  \midrule
  \multirow{8}{*}{multi-$\beta$ (all peaks)}
  & $\beta_{1}$ & $0.89\pm 0.62$ \\
  & $\beta_{2}$ & $0.49\pm 0.07$ \\
  & $\beta_{3}$ & $0.59\pm 0.19$ \\
  & $r_{c1}$ & $(327.41 \pm 158.08) \:\textrm{kpc}$ \\
  & $r_{c2}$ & $(128.51 \pm 17.10) \:\textrm{kpc}$ \\
  & $r_{c3}$ & $(42.51 \pm 13.33) \:\textrm{kpc}$ \\
  & $I_{01}$ & $75.53\pm 3.56$ \\
  & $I_{02}$ & $119.22\pm 6.10$ \\
  & $I_{03}$ & $69.22\pm 5.16$ \\
  \bottomrule
  \end{tabular}

\label{paran}
\end{table}
\subsubsection{The Convergence $\kappa$-Map for the Train Wreck Cluster}

\begin{figure}[H]
\centering
	\begin{subfigure}[b]{0.48\textwidth}
	\centering 
	\includegraphics[width=\textwidth]{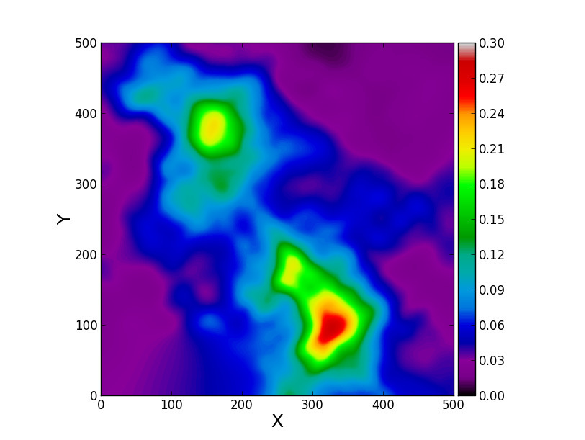}
	\caption{$\kappa$-map data from Jee et al.}
	\label{3dk}
	\end{subfigure}
	\begin{subfigure}[b]{0.48\textwidth}
	\centering 
	\includegraphics[width=\textwidth]{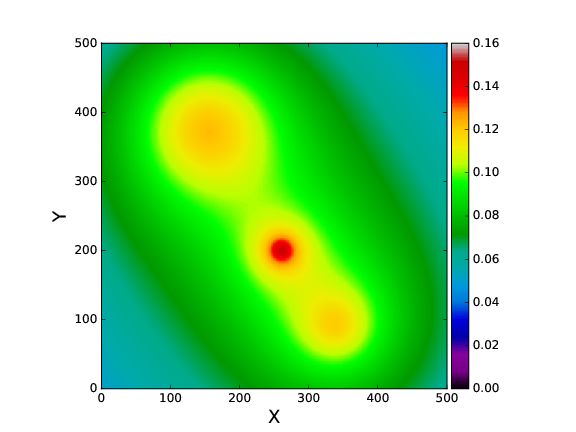}
	\caption{Density map for STVG prediction}
	\label{cjd}
	\end{subfigure}
	\vspace{-12pt}

	 \caption{A520 $\kappa$-map. The first plot is the data, whereas the second plot is the STVG (MOG) prediction.}
	\label{k_mapp1}
\end{figure}

Using the parameters for the photon spectra, the $\kappa$-map was reproduced. The full list of parameters used for the $\kappa$-map is reflected in Table~\ref{paran}. The density plot (Figure~\ref{k_mapp1}a) is a $500\times 500~\textrm{px}^2$ image. The~distance resolution for the $\kappa$-map data is computed to be $2.8~\textrm{kpc}/\textrm{px}$. This was computed by using the distance between peaks P3 and P3$'$ of Figure~\ref{k_mapp1}b. This is a distance of 1$'$ (given in~\cite{jeetal}). Figure~\ref{k_mapp1} (top and bottom) is the $\kappa$-map prediction of STVG. These results are obtained using the parameters of the fitting of the photon map given in Table~\ref{paran}.

\subsubsection{Including the Galaxies}

\begin{figure}[H]
	\centering 
	\includegraphics[width=0.62\columnwidth]{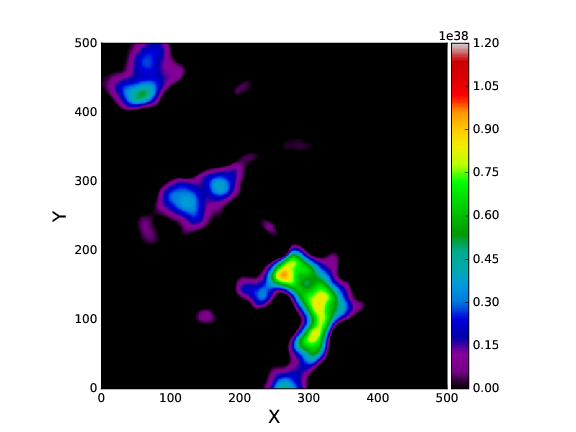}
	 \caption{A520 galaxy inclusion. The galaxies explain the regions shown. These regions are offset from the gas.}
	\label{k_gal} 
\end{figure}

The offsets of the lensing peaks for the Train Wreck Cluster are due to the ordinary baryon matter in the galaxies. Just as with the Bullet Cluster, the baryon matter in the galaxies at the the sides of the cluster merger shifts the lensing peaks from the location of the gas, producing the offset. The enhancement of the lensing peaks is due to the larger gravitational coupling produced by $G=G_N(1+\alpha)$ on large distance scales. The galaxy results are shown in Figure~\ref{k_gal}. The results amount to a galaxy contribution of $\textrm{M} = 7.0\times 10^{12}~\textrm{M}_{\odot}$. This is about $25.7\%$ of the total mass of the system. This~is neglecting the negative mass contributions, for again, such contributions are not physical. Including these contributions gives a galaxy contribution of $-88.6\%$. Again, these negative mass contributions stem from the fact that the predictions come from a zeroth order approximation, which is much more uniform than the data. The result is that some regions will have a relatively higher $\kappa$ than the same region of the data. Recall that we are looking at the difference between the data and prediction; hence, the resulting galaxy mass contribution for such regions is negative. This is an overall negative mass contribution, which is meaningless. The remaining gravity comes from the gas whose total mass was estimated to be about $\textrm{M} = 2.77\times 10^{13}~\textrm{M}_{\odot}$.

\section{Summary}\label{disc}

We summarize the Bullet Cluster results, as well as the A520 results in Table~\ref{datasum}. In both systems, the galaxy contributions are believed to be large enough to have a significant effect on the systems. The galaxy contributions were estimated to be about $14.1\%$ for the BC and $25.7\%$ for A520, and the remaining physics of both systems is explained by the gas.

The offset of the lensing peaks from the gas in both systems is due to the baryons in the galaxies, which are off to the sides, shifting the positions of the lensing peaks away from the gas. The heights of the lensing peaks are determined by the size of the parameter $\alpha$. Table~\ref{datasum} summarizes the overall results for both systems. We see that the A520 gas mass is smaller than that of the Bullet Cluster by almost a factor of 10, making the size of $\alpha$ for A520 about 10-times bigger than the alpha for the Bullet Cluster.
\begin{table}[H]
\centering
\caption{Summary of results for the Bullet Cluster and the two datasets for A520.}

  \begin{tabular}{ccccc}
  \toprule
 \textbf{ Dataset} & \boldmath{$\alpha$} & \boldmath{$\textrm{M}_{g}(10^{13}~\textrm{M}_{\odot})$} & \boldmath{$\textrm{M}_T(10^{14}~\textrm{M}_{\odot})$} &\boldmath{ $\%$}\\
  \midrule
  BC & $2.32$ & $3.1$ & $2.2$ & $14.1$ \\
 	A520 & $34.20$ & $0.7$ & $0.277$ & $25.7$\\ \bottomrule
  \end{tabular}

\label{datasum}
\end{table}

\section{Conclusions}\label{con}

Starting with the King-$\beta$ model, STVG is used to explain the Bullet Cluster and Train Wreck Cluster mergers. In the case of the Train Wreck Cluster, there are two conflicting situations, one with a lensing peak centred at the gas and the other with no lensing peak on the gas. The aim of this study is not to resolve this controversy, but to demonstrate that STVG can explain the Train Wreck Cluster without dark matter if the central peak is indeed present. This contrasts with dark matter models, which, if the central peak is indeed present, will need a modification of the standard cold dark matter model (which may include interacting dark matter) to explain it.

Recall that we take the difference between the $\kappa$-map data and the STVG prediction to estimate the amount of galaxies in the systems and the baryon matter they contain. The offset of the lensing peaks from the gas is due to the enhanced strength of gravitation (with $G=G_N(1+\alpha)$), acting on the source of ordinary matter in the galaxies at the sides of the merger. The galaxy estimate for the Bullet Cluster is $14.1\%$ and $25.7\%$ for the Train Wreck Cluster. The existence of the central lensing peak located on the gas is explained in STVG without postulating dark matter. The peak is naturally incorporated in STVG, as all one needs is the baryonic matter to explain the peak.

A future possible study is to develop a simulation model to further investigate the offset of the lensing peaks from the central gas, due to the galaxies and enhanced gravity. This simulation would approach the fitting of the gas by using an elliptical $\beta$ model instead of the spherical $\beta$ model (which was used in this study). Another study is to perform a cluster dynamics survey to see if STVG can explain other galaxy cluster mergers.

We believe that this study has successfully demonstrated that STVG or MOG can explain the mergers of the Bullet Cluster and the Train Wreck Cluster without dark matter.

\vspace{6pt}

\acknowledgments{We thank Myungkook Jee for providing data for Abell 520 and Andisheh Mahdavi for helpful discussions regarding Abell 520. We also thank David Harvey for providing further data and for useful discussions. In addition, we thank Viktor Toth, Martin Green and Joel Brownstein for helpful suggestions and discussions.}

\authorcontributions{Norman Israel played a major role in the analysis of the data in the context of the theory. John Moffat also played a major role in the data analysis. Both authors contributed equally to the writing of the manuscript.}

\conflictsofinterest{The authors declare no conflict of interest.} 


\appendixtitles{yes} 
\appendixsections{multiple}
\appendix

\section{Computing the $\Sigma$-Map for a System with \textit{n} Peaks}
\label{sigmap}
In this section, we will derive the equation for the $\Sigma$-map fitting. We start from the King-$\beta$ model and proceed as was done in~\cite{BrownsteinMoffat2}. This will be done first as a reproduction of the derivation of~\cite{BrownsteinMoffat2} for consistency, and then, we will extend the result to include the sub-cluster. We will then do a generalization to include a system with $n$ peaks.

Using the King-$\beta$ model, we treat a system as if it were an isothermal gas sphere. Starting with the BC, which has two peaks, we have the density profile:
\begin{equation}
\rho(x,y,z) = \sum_{i=1}^{n}\rho_{i}(0)\left[1+\frac{x^2_i+y^2_i+z^2_i}{r^2_{ci}}\right]^{-3\beta_{i}/2},
\label{Adens}
\end{equation}
where $\rho_{i}(0)$ is the central density corresponding to the i-th peak. This density profile will now be projected along the line of site:
\begin{equation}
\Sigma(x,y) = \int_{-z_{\textrm{out}}}^{z_{\textrm{out}}}\rho(x,y,z)dz.
\label{Aproj}
\end{equation}

We substitute \eqref{Adens} into \eqref{Aproj}, giving:
\begin{equation}
\Sigma(x,y) = \sum_{i=1}^{n}\rho_{i}(0)\int_{-z_{\textrm{out}}}^{z_{\textrm{out}}}\left[1+\frac{x^2_i+y^2_i+z^2_i}{r^2_{ci}}\right]^{-3\beta_{i}/2}dz.
\label{APrj}
\end{equation}

We do a $u$ substitution to solve the above integral \eqref{APrj}, with:
\begin{equation}
u^2 = 1+\frac{x^2+y^2}{r^2_c}
\label{usub}
\end{equation}
giving: 
\begin{align}
\Sigma(x,y) &= \sum_{i=1}^{n}\rho_{i}(0)\int_{-z_{\textrm{out}}}^{z_{\textrm{out}}}\left[u^2_i+\frac{z^2}{r^2_{ci}}\right]^{-3\beta_{i}/2}dz \nonumber \\
&= \sum_{i=1}^{n}\frac{\rho_{i}(0)}{u^{3\beta_i}_i}\int_{-z_{\textrm{out}}}^{z_{\textrm{out}}}\left[1+\left(\frac{z}{u_{i}r_{ci}}\right)^2\right]^{-3\beta_{i}/2}dz \nonumber \\
&= \sum_{i=1}^{n}2\frac{\rho_{i}(0)}{u^{3\beta_i}_i}z_{\textrm{out}}F\left(\left[\frac{1}{2},\frac{3}{2}\beta_i\right],\left[\frac{3}{2}\right],-\left(\frac{z_{\textrm{out}}}{u_{i}r_{ci}}\right)^2\right)\nonumber \\
&= \sum_{i=1}^{n}2\rho_{i}(0)z_{\textrm{out}}\left(1+\frac{x^2_i+y^2_i}{r^2_{ci}}\right)^{-3\beta_i/2}
F\left(\left[\frac{1}{2},\frac{3}{2}\beta_i\right],\left[\frac{3}{2}\right],-\left(\frac{z_{\textrm{out}}}{u_{i}r_{ci}}\right)^2\right).\nonumber 
\end{align}

Here, \emph{F} is the Gauss hypergeometric function. The full definition of this function in terms of Gamma functions is:
\begingroup\makeatletter\def\f@size{9}\check@mathfonts
\def\maketag@@@#1{\hbox{\m@th\normalsize \normalfont#1}}%
\begin{multline}
F\left(\left[\frac{1}{2},\frac{3}{2}\beta_i\right],\left[\frac{3}{2}\right],-\left(\frac{z_{\textrm{out}}}{u_ir_{ci}}\right)^2\right) = \left(1-\frac{z_{\textrm{out}}}{u_ir_{ci}}\right)^{-\frac{1}{2}}\times
\frac{\sqrt{\pi}\Gamma \left(\frac{3\beta_i-1}{2}\right)}{2\Gamma \left(\frac{3\beta_i}{2}\right)} + \left(1-\frac{z_{\textrm{out}}}{u_ir_{ci}}\right)^{-\frac{3\beta_i}{2}}
\frac{\Gamma \left(\frac{1-3\beta_i}{2}\right)}{2\Gamma \left(\frac{3(1-\beta_i)}{2}\right)}.
\end{multline}
\endgroup

We will next perform a series of approximations to simplify the above expression. The following are the approximations we will make.
\begin{enumerate}[leftmargin=*,labelsep=5mm]
\item When on the \emph{i}-th peak, $\Sigma_i(x_i,y_i)>>\Sigma_{i+1}(x_{i+1},y_{i+1})$
\item When on the (\emph{i}+1)-th peak, $\Sigma_{i+1}(x_{i+1},y_{i+1})>>\Sigma_i(x_i,y_i)$
\item $z_{\textrm{out}}>>r_{ci}$
\item $\beta_i >> 1/3$
\end{enumerate}
then we can derive an expression for the $\Sigma$-map that can apply to multiple peaks.

Using Approximations (i) and (ii), we have for both peaks:
\begin{equation}
\Sigma_{i}(0,0) = 2\rho_{i}(0)z_{\textrm{out}}F\left(\left[\frac{1}{2},\frac{3}{2}\beta_{1,2}\right],\left[\frac{3}{2}\right],-\left(\frac{z_{\textrm{out}}}{u_{i}r_{ci}}\right)^2\right).
\label{sgm0}
\end{equation}

We use this to get: 
\begin{equation}
\Sigma(x,y) = \sum_{i=1}^{n}\Sigma_{0i}\left(1+\frac{x^2_i+y^2_i}{r^2_{ci}}\right)^{-3\beta_i/2} \\ 
F\left(\left[\frac{1}{2},\frac{3}{2}\beta_i\right],\left[\frac{3}{2}\right],-\left(\frac{z_{\textrm{out}}}{u_{i}r_{ci}}\right)^2\right).
\label{sigma_hyp}
\end{equation}

Using Assumptions (iii) and (iv), we have the Gauss hypergeometric functions simplifying to Gamma functions allowing the approximation for \eqref{sigma_hyp} to be:
\begin{equation}
\Sigma_{0i} = \sqrt{\pi}\rho_{0i}r_{ci}\frac{\Gamma(\frac{3\beta_i-1}{2})}{\Gamma(\frac{3\beta_i}{2})}.
\end{equation}

Hence, from Assumptions (i)--(iv), we have:
\begin{equation}
\Sigma(x,y) = \sum_{i=1}^{n}\Sigma_{0i}\left(1+\frac{x^2_i+y^2_i}{r^2_{ci}}\right)^{-(3\beta_i-1)/2},
\end{equation}
which is what is used to fit the $\Sigma$-map data. In order to get the $\Sigma$ map for BC and A520, we~substitute $n=2$ for BC and $n=3$ for A520.

\section*{References}
\vspace{-26pt}


\end{document}